\documentclass[%
 aip,
 amsmath,amssymb,
 reprint,%
]{revtex4-1}
\usepackage{calrsfs}
\usepackage{graphicx}% Include figure files
\usepackage{dcolumn}% Align table columns on decimal point
\usepackage{bm}% bold math
\usepackage{siunitx}
\usepackage[utf8]{inputenc}
\usepackage[T1]{fontenc}
\usepackage{mathptmx}
\usepackage{etoolbox}
\usepackage{dsfont}
\usepackage{amsmath,amssymb}
\newcommand*{\citen}[1]{%
  \begingroup
    \romannumeral-`\x % remove space at the beginning of \setcitestyle
    \setcitestyle{numbers}%
    \cite{#1}%
  \endgroup
}

\makeatletter
\def\@email#1#2{%
 \endgroup
 \patchcmd{\titleblock@produce}
  {\frontmatter@RRAPformat}
  {\frontmatter@RRAPformat{\produce@RRAP{*#1\href{mailto:#2}{#2}}}\frontmatter@RRAPformat}
  {}{}
}%
\makeatother
\begin{document}
\preprint{AIP/123-QED}

\title{Chiral Vibrational Modes in Small Molecules}
% Force line breaks with \\
\author{Jichen Feng}
\affiliation{ 
Department of Chemistry, University of Pennsylvania, Philadelphia, Pennsylvania 19104, USA}
\email{jcfeng@sas.upenn.edu}
\affiliation{Department of Chemistry, Princeton University, Princeton, New Jersey 
 08544, USA}%Lines break automatically or can be forced with \\
\author{Ethan Abraham}%
\affiliation{ 
Department of Physics, University of Pennsylvania, Philadelphia, Pennsylvania 19104, USA}
\email{ethana@mit.edu}
\author{Joseph Subotnik}
\affiliation{ 
Department of Chemistry, University of Pennsylvania, Philadelphia, Pennsylvania 19104, USA}
\affiliation{Department of Chemistry, Princeton University, Princeton, New Jersey 08544, USA}
\email{subotnik@princeton.edu}
\author{Abraham Nitzan}%
\affiliation{ 
Department of Chemistry, University of Pennsylvania, Philadelphia, Pennsylvania 19104, USA}
\affiliation{ 
School of Chemistry, Tel Aviv University, Tel Aviv 69978, Israel
}
\email{anitzan@sas.upenn.edu}

\date{\today}% It is always \today, today,
\begin{abstract}

The development of quantitative methods for characterizing molecular chirality can provide an important tool for studying chirality induced phenomena in molecular systems. Significant progress has been made in recent years toward understanding the chirality of molecular normal vibrational modes, mostly focusing on vibrations of helical molecular structures. In the present study, we examine the applicability two methodologies previously used for helical structures for the quantification of the chirality of molecular normal modes across a range of small, not necessarily helical, molecules. The first approach involves the application of the Continuous Chirality Measure (CCM) to each normal mode by associating the mode with a structure formed by imposing the corresponding motion about a common origin. The second approach assigns to each normal mode a pseudoscalar defined as the product of atomic linear and angular momentum summed over all atoms. In particular, using the CCM also as a measure of the chirality of the underlying molecular structure, we establish the existence of correlation between the chirality of molecular normal modes and that of the underlying molecular structure. Furthermore, we find that normal modes associated with different frequency ranges of the molecular vibrational spectrum exhibit distinct handedness behavior. 
\end{abstract}
\maketitle

\section{Introduction}
Molecular chirality has long been studied with regard to its chemical and optical implications, and developing techniques for characterizing and separating enantiomers has been of paramount importance to chemistry for many years.\cite{hicks2002chirality}

Considerable attention has been recently focused on a slightly different problem, namely the role molecular chirality in promoting spin selective electron transport through chiral molecular assemblies. The underlying mechanism for this Chiral Induced Spin Selectivity (CISS) effect is still a subject of ongoing discussion, and some theoretical calculations \cite{fransson2019theory,yan2023locking,yan2023locking2,yan2024lockingreview,abe2008theory,abe2013theory} and experimental observations \cite{naaman2012exp,naaman2020exp,waldek2024review} suggest the possible involvement of chiral phonons, often associated with normal modes that carry atomic angular momenta. The existence and implications of such atomic motions has been subject of several recent studies, mostly in condensed matter physics,\cite{zhang2014theory,zhang2015theory,zhang2019theory,zhu2018phononexp,chen2018review,chen2022diodetheory} where behaviors associated with correlation (locking) of atomic linear and angular momenta were observed. 

Extending these solid state considerations observed to molecular structures in which (pseudo) linear momentum is not a good quantum number is not straightforward. Still, studies of chirality-induced phenomena in molecular systems are expected to be facilitated by quantitative characterization of their chirality. While no unique measure for quantifying chirality can be formulated, \cite{Kamien} several useful measures have been studied. We have recently applied two methodologies to quantify the chirality of molecular normal modes.\cite{ethan2023quantifying,ethan2024benchmark} The first approach involves the application of the Continuous Chirality Measure (CCM) \cite{avnir1995CCM,avnir2005CCM,avnir2008CCM} to any normal mode by associating the mode with a structure formed by imposing the corresponding atomic motions on a common origin. The second  assigns to each normal mode a pseudoscalar defined as the sum over all atoms of products of of the components along some characteristic molecular axis $z$ of the atomic linear and angular momentum vectors. Our analysis was based primarily on numerical experiments done on double-helical structure with controlled amount of twist.\cite{ethan2023quantifying,ethan2023chain}

Here we extend these studies to relatively small, non-helical molecules, and show that the concepts and correlations observed for helical structures apply for such molecules as well. To this end we have performed calculations similar to those described in a work done by Abraham \textit{et al.} \cite{ethan2023chain,ethan2023quantifying,ethan2024benchmark} on the set of small molecular structures listed in Table 1. The molecules chosen for this study  contain fewer than 14 atoms based on hydrocarbon and hydrosilicon compounds sometimes with nitrogen and oxygen substitutions. We find, that the chirality measures for vibrational normal modes examined in our earlier studies of helical molecular structures are relevant also for these non-helical small molecules. In particular, using the CCM also as a measure of the chirality of the underlying molecular structure, we establish the existence of correlation between the chirality of molecular normal modes and that of the underlying molecular structure. Furthermore, we find that normal modes associated with different frequency ranges of the molecular vibrational spectrum exhibit distinct handedness behavior.
\begin{table}
    \centering
    \begin{ruledtabular}
    \begin{tabular}{lccc}
    Molecule & \# of Atoms& $C_S$ & $z$ axis\\
\hline
        $\mathrm{H_3C-CH_3}$ & 8 & 0 & C-C\\
        $\mathrm{(OH)_2P-P(OH)_2}$ & 10 & 0 & P-P\\
        $\mathrm{H_3Si-SiH_3}$ & 8 & 0 & Si-Si\\
        $\mathrm{Na(H_2Si)-(SiH_2)Na}$ & 8 &  0& Si-Si\\
        $\mathrm{(OH)(H_2C)-(CH_2)(OH)}$ & 10 & 0 & N/A\\
        $\mathrm{C_6H_6}$ & 12 & 0 & N/A\\
        $\mathrm{CH_3ON}$ & 6 & 0 & N/A\\
        $\mathrm{(HO)Li_2Si-SiLi(OH)_2}$ & 10 & 0.00287 & Si-Si\\
        $\mathrm{C_2H_4O_3}$ & 9 & 0.01189 & N/A\\
        $\mathrm{H_2N-NH_2}$ & 6 & 0.01699 & N-N\\
        $\mathrm{CH_5ON}$ & 8 & 0.02066 & N/A \\
        $\mathrm{(H_2N)_2P-P(NH_2)_2}$ & 14 & 0.02282 & P-P \\
        $\mathrm{(CHO)_2C-C(CHO)_2}$ & 14 & 0.02735 & C-C \\
        $\mathrm{(H_2P)_2P-P(PH_2)_2}$ & 14 &0.03136& P-P \\
        $\mathrm{H_2(H_3C)Si-Si(CH_3)H_2}$ & 14 & 0.03209 & N/A \\
        $\mathrm{H_2P-PH_2}$ & 6 &0.0417& P-P \\
        $\mathrm{HLi_2Si-SiLi_2H}$ & 8 &0.04374& Si-Si \\
        $\mathrm{CH_5O_2N}$ & 9 & 0.046 & N/A \\
        $\mathrm{(H_2P)_2N-N(PH_2)_2}$ & 14 & 0.04984 & N-N \\
        $\mathrm{H_2(HO)Si-Si(OH)H_2}$ & 10 &0.05131& Si-Si \\
        $\mathrm{(HO)_4N-N(OH)_4}$ & 10 &0.05994& N-N \\
        $\mathrm{H(HO)_2C-C(OH)_2H}$ & 12 &0.06624& C-C \\
        $\mathrm{(HO)_2LiSi-SiLi(OH)_2}$ & 12 & 0.07541 & Si-Si \\
        $\mathrm{(NH_2)H_2C-CH_2(OH)}$ & 11 & 0.09732 & N/A \\
        $\mathrm{H_3C-CH(OH)(NH_2)~~~(S)}$ & 11 & 0.10058 & N/A \\
        $\mathrm{C_2H_8ON_2~~(S)}$ & 13 &0.11538& N/A \\
    \end{tabular}
    \caption{The molecular set used in this study. Shown are the computed CCM of the equilibrium molecular structure and the axis chosen for the calculation of the momentum pseudoscalar. N/A indicates the molecule doesn't have a proper axis for helicity calculation, thus the helicity data isn't included in all the result.}
    \end{ruledtabular}
    \label{tab:my_label}
\end{table}

\section{\label{sec:level1}Chirality measures and their calculation  }
As stated in the introduction, the present study focuses on two chirality measures. One of them, the continuous chirality measurement (CCM),\cite{avnir1995CCM,avnir2005CCM,avnir2008CCM} has long been used to characterize the chirality of equilibrium molecular structures and has been recently generalized as a characteristic of the chirality of molecular normal modes.\cite{ethan2023quantifying,ethan2024benchmark} The other, the momentum pseudoscalar ($\mathcal{H}$) defined by Eq.~\ref{equation:helicity} below, has been recently introduced by us as a quantifier of the chirality of molecular normal modes.\cite{ethan2023quantifying,ethan2024benchmark} In what follows we briefly review these quantifiers.

\subsection{\label{sec:level2}Continuous Chirality Measure (CCM)}
The \textit{continuous chirality measure} is the result of a mathematical definition and a computational procedure that yields a distance between a given discrete structure and its nearest mirror image \cite{avnir1995CCM,avnir2005CCM,avnir2008CCM}. The mathematical expression for the CCM of the molecular equilibrium structure is:

\begin{equation}
\label{equation:CCM}
    CCM(Q)=\mathrm{min}\left\{\frac{1}{2}- \frac{ \sum_{i=1}^N \langle \mathbf{q}_i|\sigma|\mathbf{p}_i\rangle}{2\sum_{i=1}^N\langle \mathbf{q}_i|\mathbf{q}_i\rangle}\right\}
\end{equation}

Here, $\mathbf{Q}=(\mathbf{q}_1,...,\mathbf{q}_N)$ represents a molecular configuration defined by the atomic equilibrium position vectors $\mathbf{q}_i=(x_i,y_i,z_i)^T$. $\mathbf{P}=(\mathbf{p}_1,...,\mathbf{p}_N)=\mathcal{P}\mathbf{Q}$  is a similar molecular configuration obtained from $\mathbf{Q}$ by some permutation $\mathcal{P}$ of atoms from identical elements (for more details, see Appendix A).  The operator $\sigma$ denotes reflection about a given  mirror plane, and $\langle \mathbf{q}|\mathbf{p} \rangle$ is the scalar product of the two vectors $\mathbf{q}$ and $\mathbf{p}$. CCM($\mathbf{Q}$)  is obtained by minimizing the expression on the right over all such permutations and all possible choices of mirror plane.  \footnote{In some published studies the permutations $\mathbf{Q} \rightarrow \mathbf{P}$ are avoided altogether so that $\mathbf{P}$ is replaced by $\mathbf{Q}$ in Eq.~\ref{equation:CCM}. This so called convention of trivial permutations used to deal with large systems where the complexity of the calculation rapidly increases with the number of identical atoms. Here, we do not restrict ourselves to such conventions, as our system is not that large, allowing us to find the best permutation. Some other considerations associated with the applications of permutations are presented in Appendix A.} 

In what follows we denote the CCM measure of equilibrium molecular structures by $C_S$. The corresponding measure $C_M$ for a molecular vibrational mode $k$ is obtained by replacing the atomic equilibrium position vectors $\mathbf{q}_i$ by the positions determined by the corresponding normal mode displacements, $\delta \mathbf{q}_i=m_i^{-1/2}|\epsilon_{k,i}\rangle$ where $|\epsilon_{k,i}\rangle=(\epsilon_{k,i}^x,\epsilon_{k,i}^y,\epsilon_{k,i}^z)^T$ is the normalized (square root of mass-weighted) displacement of atom $i$ under normal mode $k$ which satisfies $\sum_i\langle \epsilon_{k,i}|\epsilon_{k',i}\rangle=\delta_{k,k'}$.

\subsection{\label{sec:level2}Momentum Pseudoscalar / Helicity}
The \textit{momentum pseudoscalar} associated with mode $k$, which we refer to also as \textit{helicity} $\mathcal{H}_k $ in this work, is defined as \cite{ethan2023quantifying}
\begin{equation}
\label{equation:helicity}
    \begin{aligned}
        \mathcal{H}_k=&\frac{1}{E_k}\sum_i p_{k,i}^z L_{k,i}^z\\
                  =&\frac{1}{E_k}\sum_i p_{k,i}^z (p_{k,i}^y x_i-p_{k,i}^x y_i)
    \end{aligned}
\end{equation}
where, for motion along mode $k$, $L_{k,i}^z$ and $p_{k,i}^z$  are, respectively, the $z$ component of the angular momentum of atom $i$ about a chosen molecular axis $z$, and the linear momentum of this atom along the same axis. We follow Ref. \citen{ethan2023quantifying} in dividing by the excitation energy $E_k$ of the mode $k$ in order to obtain a quantity that is intrinsic to the mode's geometry and independent of the excitation level. In particular, denoting the amplitude of mode $k$ by $ A_k $ so that the corresponding displacement of atom $i$ is $\delta \mathbf{q}_i(t)=A_k(t)m_i^{-1/2}|\epsilon_{k,i}\rangle$, the momentum of atom $i$ moving along this mode is \( \mathbf{p}_{k,i}(t) = \dot{A}_k(t)\sqrt{m_i} | \epsilon_{k,i} \rangle \) where $ A_k(t) = e^{i\omega t} A_k$. Note that $x_i$ and $y_i$ in Eq.~\ref{equation:helicity} are components not of $\delta \mathbf{q}_i(t)$  but of the vector corresponding to the equilibrium distance of atom $i$ from the molecular axis $z$. Since the mass-weighted mode coordinates are normalized to unity, we note that the mode's energy is given by $E_k=\dot{A}_k ^2,$ and so Eq. 2 can be cast in terms of the normal mode coordinates as $\mathcal{H}_k=\sum m_i\epsilon_{k,i}^z(x_i \epsilon^y_{k,i} - y_i \epsilon^x_{k,i})$.

The mode helicity $\mathcal{H}_k $  is a measure of correlation between the linear and angular momenta associated with motion under this mode. Note that $\mathbf{p}\cdot\mathbf{L}=\mathbf{p}\cdot(\mathbf{r}\times \mathbf{p})$ is zero by definition; however, the component along one axis $p^zL^z$  is an measure of such correlation that can be nonzero. The axis $z$ is chosen to represent some conceived symmetry in the molecule and in the molecular set used in the present study it as taken along a C-C or Si-Si bond. Obviously this choice is not unique and its effect on our results is discussed in the appendix. We find that this choice affects the calculated helicities but not  the correlations and trends presented below. The latter observation is consistent with the fact that the contributions $p^x_{k,i}L^x_{k,i}$, $p^y_{k,i}L^y_{k,i}$, $p^z_{k,i}L^z_{k,i}$ must be correlated because their sum must vanish. 

As discussed in Ref.~\citen{ethan2023quantifying}, we note that while the CCM is effectively a measure of the similarity of an object with its enantiomer and thus gives the same positive value for both enantiomers, the momentum pseudoscalar indeed gives opposite signs for opposite enantiomers. 

In the following section, we present results of calculations of these chirality measures for the set of molecules listed in Table~1. We emphasize the difference between these quantitative measures of chirality, which depend on a molecule's conformation, and the conventional binary R/S assignment, which depends on the molecule's connectivity. As such, we see that although most molecules in Table~1 (with the exception of the last two molecules) do not have chirality expressed in the R/S sense, many of these molecules still have small nonzero $C_S$ values for their equilibrium conformation (which by symmetry will be degenerate with an enantiomeric conformation).

These calculations serve to demonstrate the correlation between the chirality of molecular structures and that of the molecular normal modes. For some of these molecules, we have further studied the effect of imposed structural distortions that effectively twist the molecular structure about a chosen axis. This leads to an observed correlation between the twist angle and the corresponding structural and mode chiralities and makes it possible to compare the two chirality measures used in the present study.\cite{ethan2023quantifying,ethan2024benchmark,ethan2023chain}

In carrying out these calculations, the molecular structure and normal modes were evaluated using density functional theory (DFT) with the B3LYP functional using the GAMESS software. The twisting procedure was done by relaxing the molecular structure to the minimum energy configuration under the constraint by a given dihedral angle. The modes with twist were obtained by vibrational mode calculation without constraints but from the twisted structure. Note that the convexity of the energy surface isn't guaranteed in this procedure as the restricted structure is not the potential minimum; that being said, we didn't encounter such a problem when performing calculations for the twisted molecules as we found all the modes frequencies to be real.

\begin{figure}
\includegraphics[width=240pt]{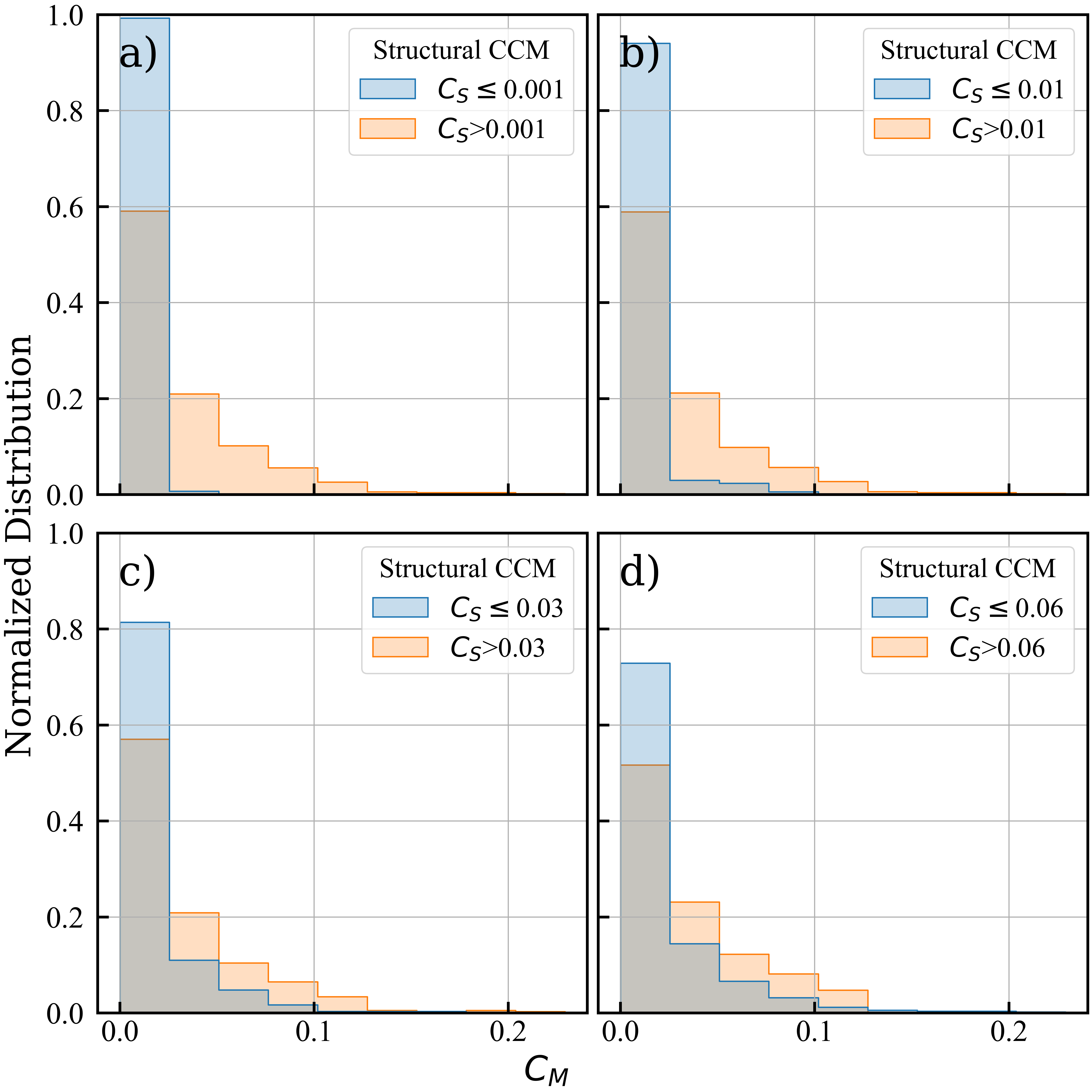}
\caption{Distribution of CCM of mode ($C_M$) values displayed shown for chiral (red, large $C_S$ value) and non-chiral (blue, small $C_S$ value) molecular structures. The threshold for characterizing the molecule as chiral was taken 0.001, 0.01, 0.03 and 0.06 in panels (a),(b),(c) and (d), respectively.}
\label{fig:CCM_and_CCM}
\end{figure}
\begin{figure}
\includegraphics[width=240pt]{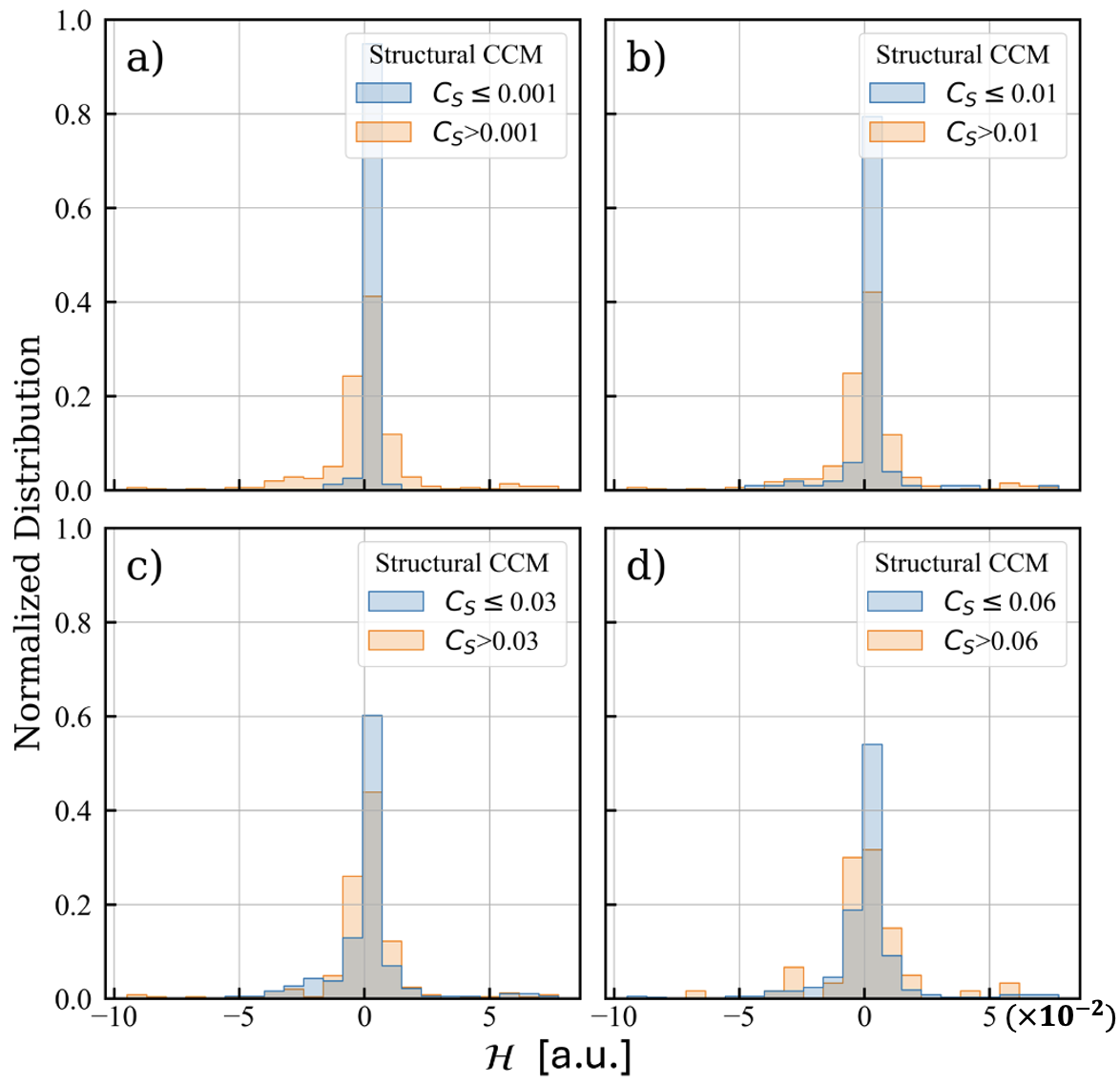}
\caption{ Same as Fig. \ref{fig:CCM_and_CCM}, but for the mode helicity values instead of the CCM. Helicity is reported in an arbitrary unit that is equivalent to $8.8 \times 10^{-38}$ kg $\cdot$ m, and also note the scaling of the $y$ axis.}
\label{fig:CCM_and_Helicity}
\end{figure}

\begin{figure}
\includegraphics[width=240pt]{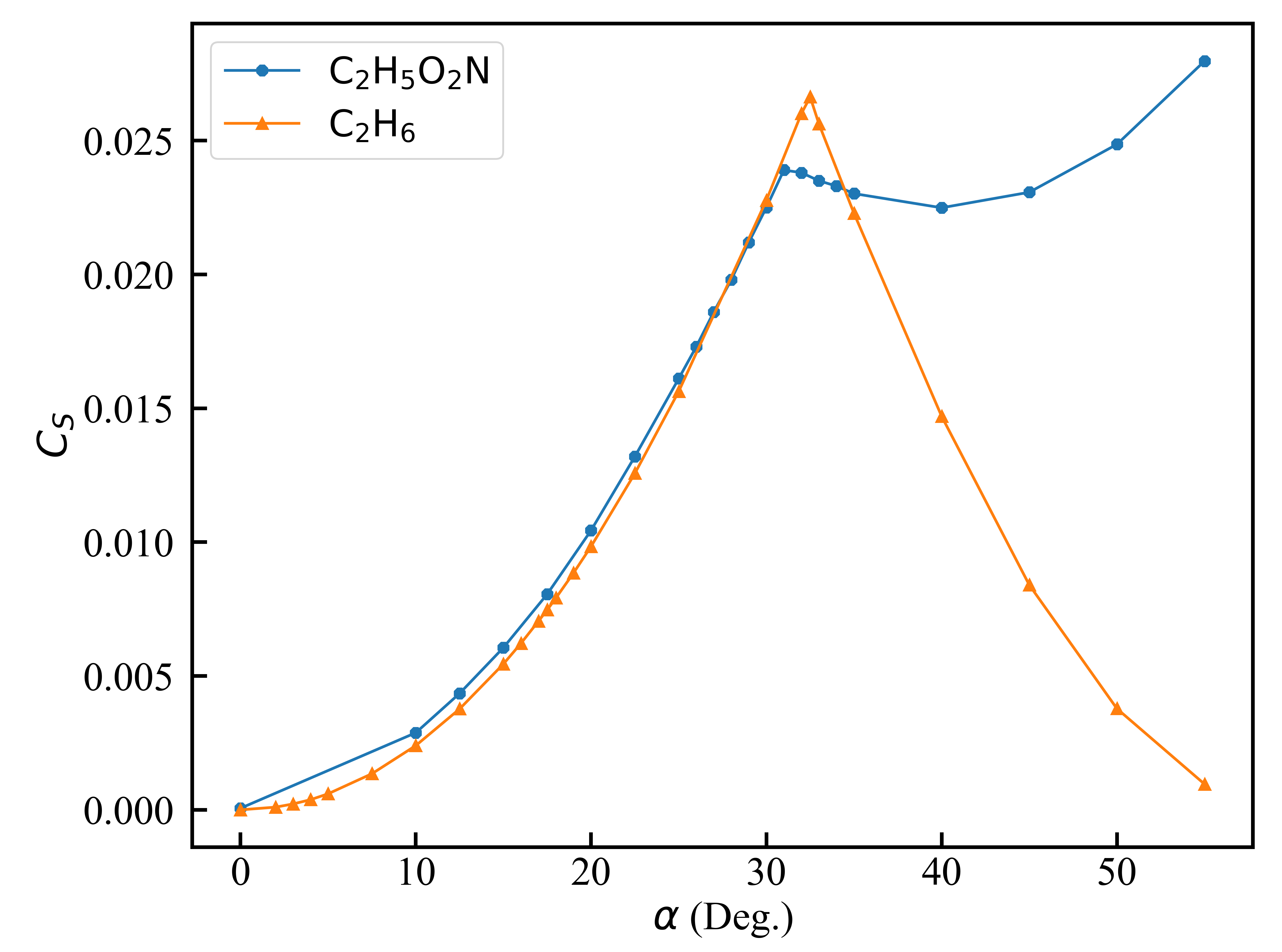}
\caption{ Correlation between $C_S$ and twist angle of ethane (orange) and glycine (blue). }
\label{fig:Str_CCM}
\end{figure}

\section{\label{sec:level1}Results}
Figures \ref{fig:CCM_and_CCM}  and \ref{fig:CCM_and_Helicity} plot the distribution of the mode CCM ($C_M$) and helicity ($\mathcal{H}$) values calculated for all modes of all chiral (red) and achiral (blue) moleules in our molecular set. For these displays, a molecule is assigned as chiral ($C_S>C_S^*$) or non-chiral character ($C_S\leq C_S^*$) according to a chosen CCM threshold $C_S^*$ which is set to 0.001, 0.01, 0.03 and 0.06 in panels (a), (b), (c), (d) respectively of both figures. Note that the $C_M$ values are always positive. In both cases we see that the vibrational modes associated with the chiral molecular structures are more likely to appear chiral than those arising from achiral molecular structures. However this correlation weakens when a more relaxed structural chirality criterion (larger $C_S^*$) is used. Also note that the correlation between $C_S$ and $\langle \mathcal{H} \rangle$ is weaker than between $C_S$ and $C_M$.

Next, we consider the behavior of our structural and mode chirality measures for molecular structures whose chiralities are varied under applied twist. The structures used in this study are obtained from the ethane ($\mathrm{C_2H_6}$) and glycine ($\mathrm{C_2H_5O_2N}$) molecules by changing the molecular forcefield with an added twist that effects a rotation by an angle $\alpha$ about the C-C bond. The twist added to ethane is right-handed around the axis, while left-handed for glycine. The equilibrium structures of the molecules are achiral, with  mirror symmetry about the molecule plane (for the ethane molecule there exists also an improper rotational symmetry about a plane perpendicular to the C-C bond). The added twist destroys these symmetries and renders the structures chiral, as shown in Fig. 3 with average mode CCM of ethane displayed in orange and that of glycine shown in blue. For the ethane-based structures, structural CCM increases quadratically (see Appendix D) as the twisting angle increases between $\alpha \in [0,30^{\circ}]$ and decreases almost symmetrically after $\alpha \sim30^{\circ}$. A similar trend is seen in the case of twisted glycine: the structural chirality increases as the twisted angle $\alpha$ increases and turns down slightly above $\alpha \sim 30^{\circ}$. In this case, however, the turnover is not symmetric as for ethane. A close look into the CCM evaluation procedure reveals that the origin of this asymmetry is the fact when $\alpha$ exceeds $30^{\circ}$, the symmetric image of the structure is obtained by a combination of reflection and permutation of hydrogen atoms.

\begin{figure}
\includegraphics[width=230pt]{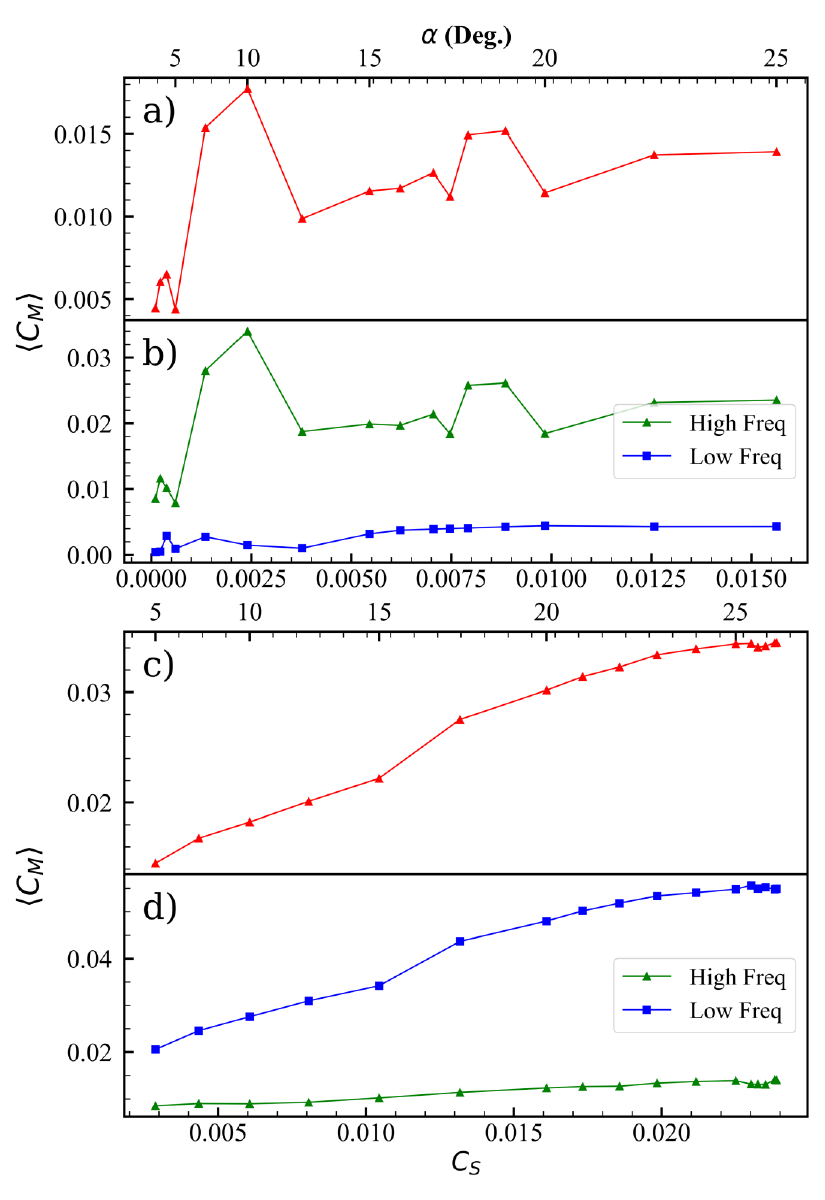}
\caption{ Correlation between $\langle C_M \rangle$, the average modes CCM, and $C_S$, the CCM of the equilibrium molecular structure plotted for twisted ethane (panels (a),(b)) and glycine (panels (c),(d)). Panels a and c show the mode CCM averaged over all molecular modes while panel (c) and (d) show the modes CCMs averaged separately over the groups of high (green triangles) and low (blue squares), where each group contain half the total number of modes.}
\label{fig:Ethane_Mode}
\end{figure}

\begin{figure}
\includegraphics[width=232pt]{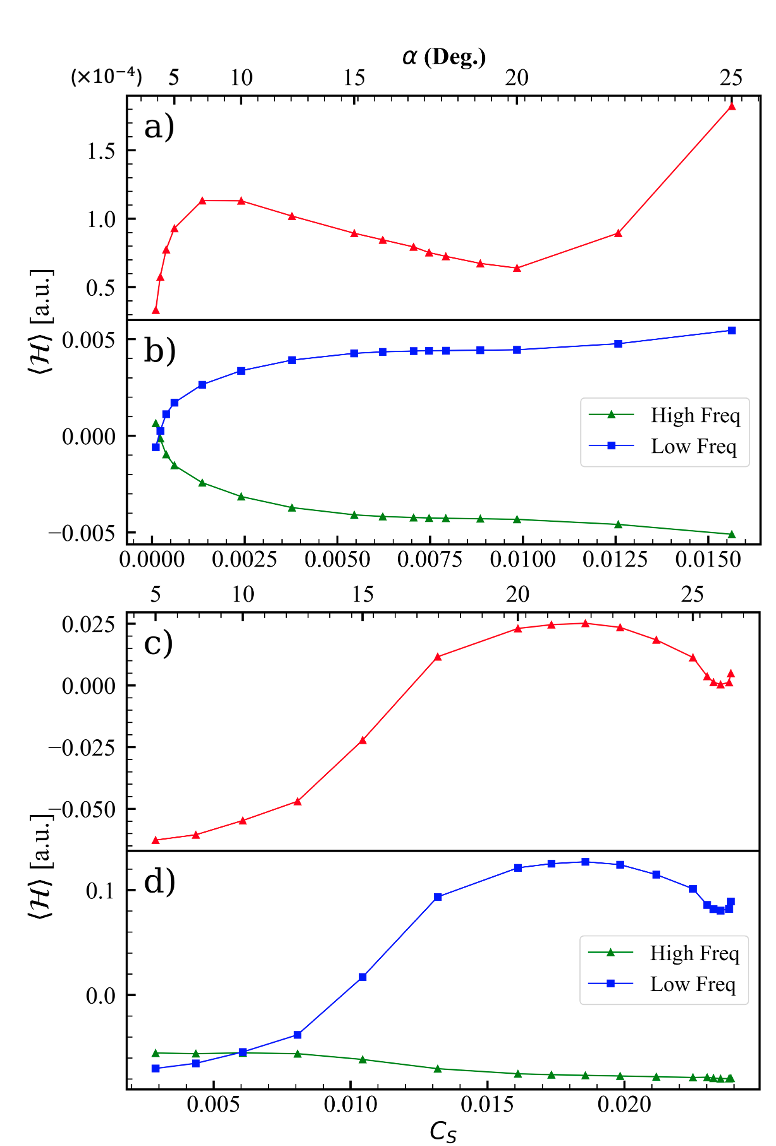}
\caption{ The same as Fig. \ref{fig:Ethane_Mode} only now with momentum pseudoscalar ($\langle \mathcal{H} \rangle$) values averaged over all modes (panel (a) and (c) and over the groups (each containing half the total number of modes) of high and low frequency modes. The calculation is performed using the helicity axis chosen as shown in Table 1. Similar plots using other directions for the helicity axis are shown in Appendix B. Helicity is reported in an arbitrary unit that is equivalent to $8.8 \times 10^{-38}$ kg $\cdot$ m, and also note the scaling of the $y$ axis.}
\label{fig:Glycine_Mode}
\end{figure}

Next consider the modes chiralities of these distorted ethane and glycine structures. Figs. \ref{fig:Ethane_Mode} and \ref{fig:Glycine_Mode} show the average mode CCM $\langle C_M \rangle$ and helicity $\langle \mathcal{H} \rangle$, respectively. Again, we show in these plots both the averages over all modes (red) as well as average over the groups (each containing half of the number of modes) of low (blue) and high (green) frequency modes. The following observations are noteworthy: (i) The average CCM of mode for both ethane and glycine is very small when $C_S=0$ (the original undistorted structures) and increases as the $C_S$ increase with distortion. (ii) Similarly, the average mode helicity $\langle \mathcal{H} \rangle$ is close to zero for $C_S=0$ and increases as $C_S$ becomes larger. (iii) The correlation between $\langle \mathcal{H} \rangle$ and $C_S$ shows opposite trends for low and high frequency modes: the $\langle \mathcal{H} \rangle$ of the former trends to more positive values while the latter becomes more negative with increasing $C_S$. These trends are similar to those observed for helical structures in Ref. \citen{ethan2023quantifying}. (iv) The correlation of $\langle C_M \rangle$ and $C_S$ is weaker in the ethane based structures than in those obtained by distorting glycine, while the opposite is true for the correlation between $\mathcal{H}$ and $C_S$ of these structures, showing stronger correlation in the ethane structures (especially when shown separately for the high and low frequency groups).

\section{Conclusions}
In this paper, we have studied the chirality of modes of chiral and achiral small molecules both relative to their equilibrium structures as well as twisted configurations. The calculations pertaining to the twisted small molecules indicate that the continuous chirality measure $C_S$ is a good measure of the chirality of the molecular structure, as it consistently shows good correlations within a finite range of twists away from the symmetric structure, supporting the observations of Refs. \citen{ethan2023quantifying, ethan2024benchmark}.

Following previous studies of helical molecular structures, two measures were examined as quantifiers of the chirality of vibrational modes: one ($C_M$) is an extension of the continuous chirality measure to normal modes. The other ($\mathcal{H} $) is the mode helicity, which measures correlation between atomic angular and linear momenta. Both measures show strong correlation with the structural chirality $C_S$, showing that chiral modes are more likely to appear in chiral molecular structures. Another interesting finding is that the handedness of chiral molecular vibrations is different for low and high frequency modes. The origin of this difference, as well the possible relevance of these observations to the optical response of chiral molecules as observed in VCD spectroscopy, will be further explored in future studies. Furthermore, as discussed in Ref. \citen{ethan2023quantifying}, the opposite behavior of the helicity (there referred to as the momentum pseudoscalar) in different frequency groups is predicted to give rise to net thermal chiral motion in which the total linear and angular momentum of the atoms are correlated in chiral structures at equilibrium. It is emphasized that this is a strictly quantum effect, as such a correlation would be forbidden classically by the equipartition theorem.\cite{ethan2023quantifying} The physical significance of this effect, and its potential relationship to other vibrational angular momentum effects that have been observed, \cite{phon_ang_mom,therm_angular} will be the subject of future investigation as well.

\section{Acknowledgement}
The research of A.N. is supported by the Air Force Office of Scientific Research under award number FA9550-23-1-0368. E.A. acknowledges the support of the University of Pennsylvania GfFMUR and Vagelos Science Challenge grants, as well as the Massachusetts Institute of Technology Department of Chemistry. We thank Claudia Climent for many helpful discussions.

\setcounter{figure}{0}
\renewcommand{\figurename}{FIG.}
\renewcommand{\thefigure}{S\arabic{figure}}

\setcounter{equation}{0}
\renewcommand{\theequation}{S\arabic{equation}}

\appendix
\section{Variations on the permutation procedure in CCM calculations}
\begin{figure}
\includegraphics[width=240pt]{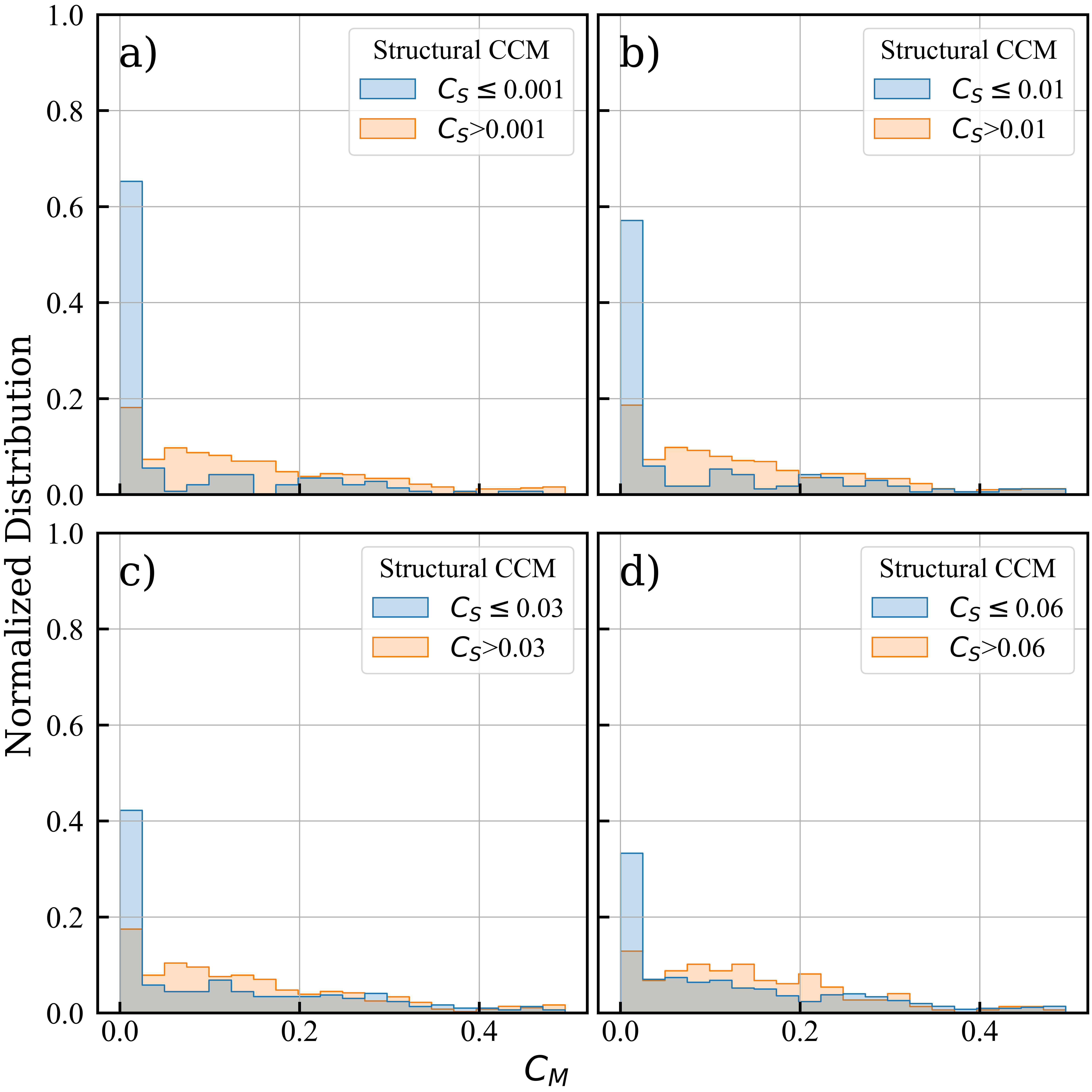}
\caption{The same as Fig. 1. The permutation of the $C_M$ is fixed to be the same as the permutation from $C_S$ calculation.}
\label{fig:CCM_and_CCM_1}
\end{figure}

As alluded to in the main text, in the CCM literature different approaches are sometime taken for applying permutations on identical atoms when evaluating the CCM from Eq.~1. In the present study we have examined the robustness of our observations to several such choices.

First, in both the $C_M$ and $C_S$ calculations shown in the main text, we have minimized the right hand side of Eq.~1 over all permutations that permute atoms of the same element type. In such a way, we are guaranteed to treat all atoms of the same element type in the most balanced fashion. Alternatively, we could have use the same permutation that minimizes $C_S$ also for evaluating $C_M$. Results obtained by using this procedure are shown in Fig.~\ref{fig:CCM_and_CCM_1}. The distribution of the mode CCMs appears to be slightly broader (and more spread out), but overall the same trend remains.

Secondly, in our previous work\cite{ethan2023chain,ethan2023quantifying,ethan2024benchmark}, we discussed the necessity of applying a notion of connectivity to structural CCM calculations. When connectivity is considered, we can distinguish identical elements occupying chemically inequivalent sites and therefore ignore any permutation of such atoms; otherwise we would by default consider all permutations that switches atoms of same element. This approach is computationally cheaper as it eliminates several possible permutations and is arguably more physically sensible.

Within the present study we have tested both approaches taking into account or ignoring connectivity when applying permutations during the structural CCM calculations. For all the small molecules examined (at their equilibrium structures, i.e. without twist), the results indicated that the structural CCM appears to be the same in both approaches. This is likely because the small, high-symmetry molecules examined are such that the cases of identical yet chemically equivalent atoms are few. An exception we found occurred when twisting glycine beyond $\sim 30^\circ$; here, the best permutation did in in fact match chemically inequivalent hydrogen atoms, which must partially explain the turning point of the structural CCM vs twisting angle for glycine at this angle as shown in Fig.~\ref{fig:Str_CCM}.

%As in the previous calculation for the $C_M$, we always choose the permutation that gives the smallest $C_M$. Here we would like to discuss the physical meaning of choosing such a permutation. With such a choice of permutation, we are treating all  atoms of the same element in an identical fashion.  XXX What does this mean?  As we only interested in their displacement and ignored their equilibrium positions. XXX

%We could also choose another permutation for $C_M$ which is the same permutation as the permutation that gives the smallest $C_S$. With such permutation, we paired the atoms that are paired during the structural CCM calculation which is more likely to be the mirrored atoms. The result of such calculation is giving a similar trend as the previous method, despite the CCM of modes are more spread. This indicates that the paring of the $C_M$ will not give the best permutation for mode calculation, but the trend remains.

\section{Dependence of the momentum pseudoscalar on the choice of axis}
\begin{figure}[hbt!]
\includegraphics[width=240pt]{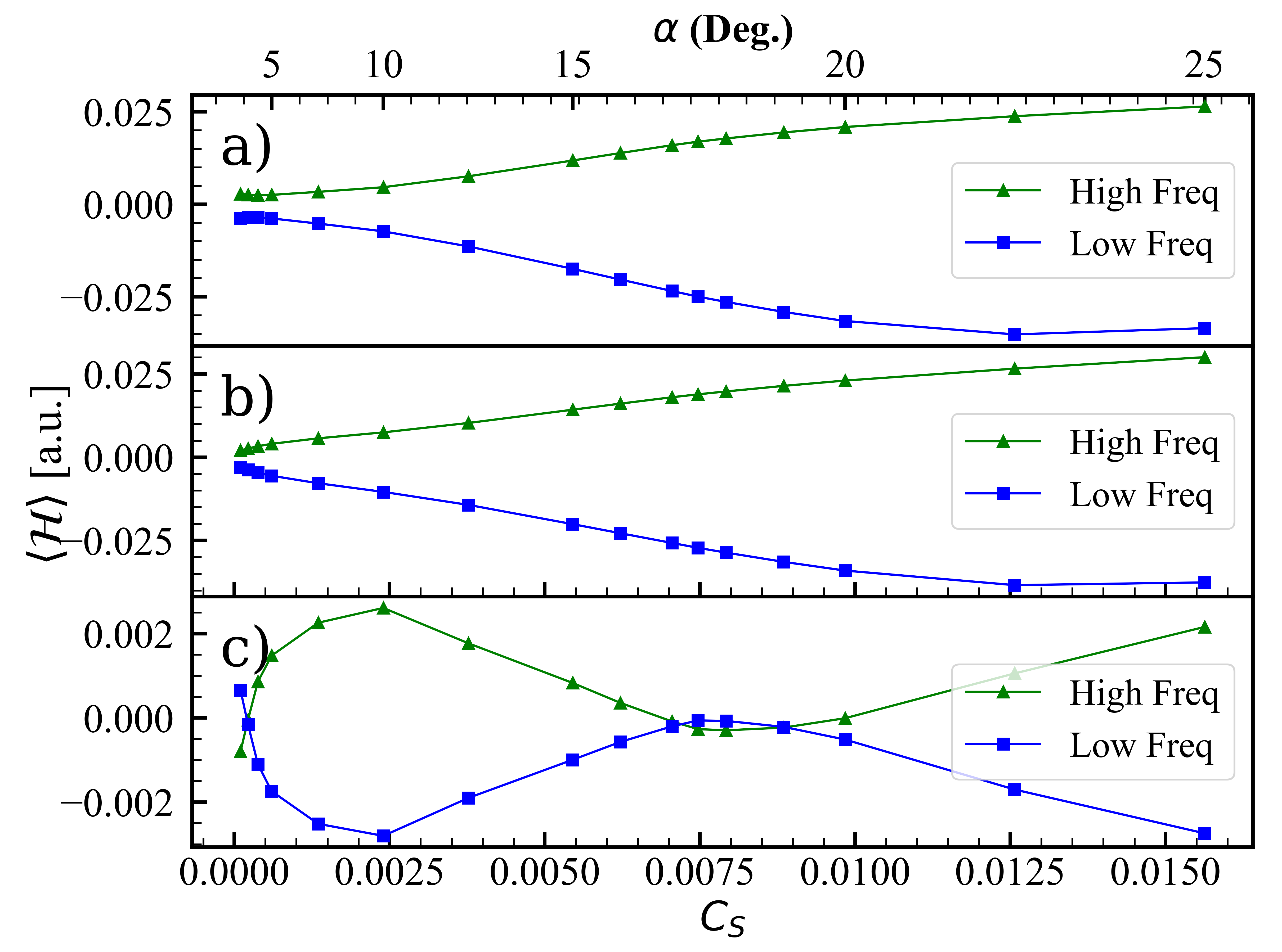}
\caption{ Rotating the axis of the momentum pseudoscalar of ethane around $x$ axis by (a) $30^\circ$ (b) $60^\circ$ (c) $90^\circ$ (Originally along $z$ axis) Helicity is reported in an arbitrary unit that is equivalent to $8.8 \times 10^{-38}$ kg $\cdot$ m, and also note the scaling of the $y$ axis.}
\label{fig:C2H6_X}
\end{figure}

\begin{figure}[hbt!]
\includegraphics[width=240pt]{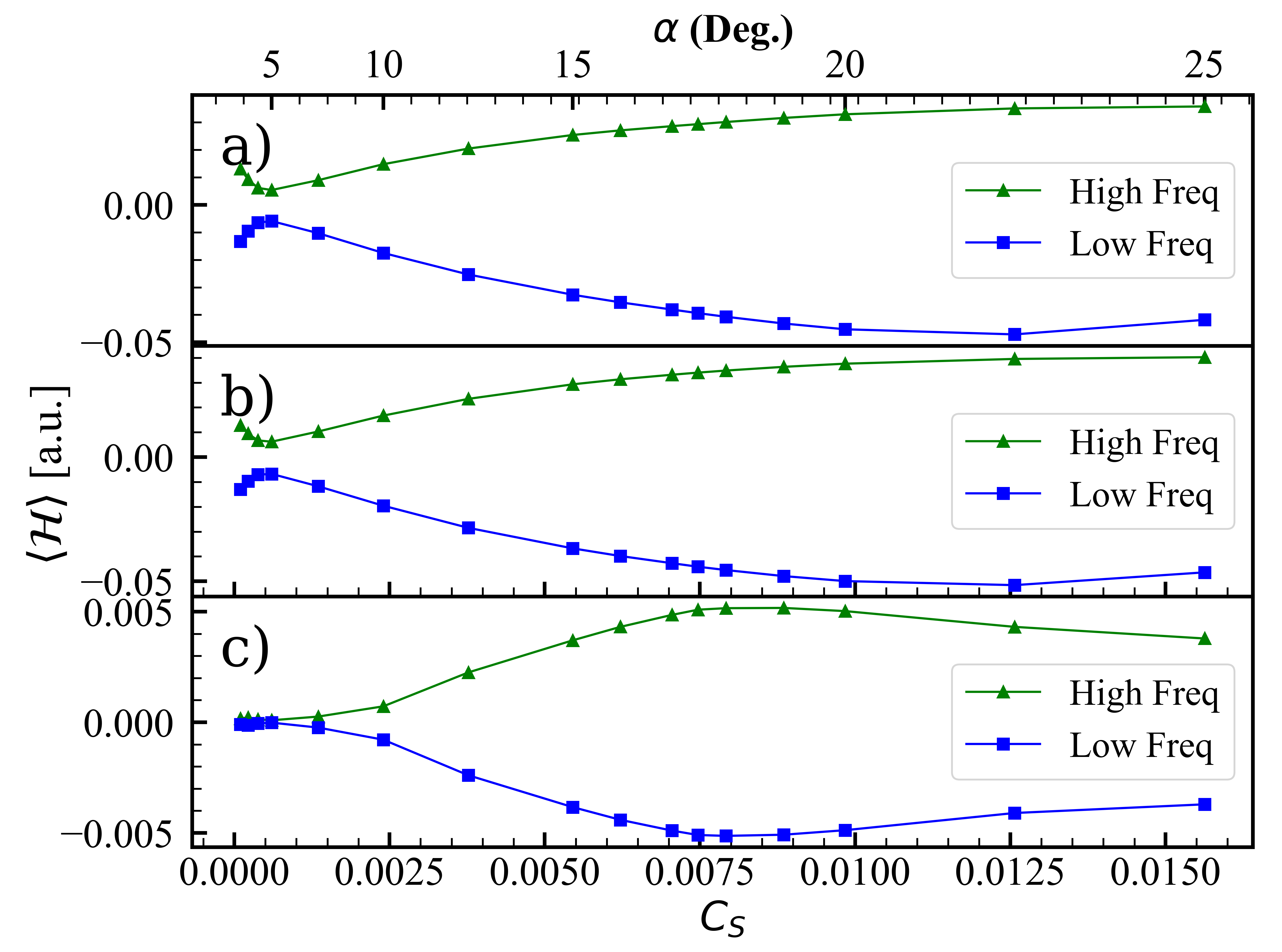}
\caption{ Rotating the axis of the momentum pseudoscalar of ethane around $y$ axis by (a) $30^\circ$ (b) $60^\circ$ (c) $90^\circ$ (Originally along $z$ axis) Helicity is reported in an arbitrary unit that is equivalent to $8.8 \times 10^{-38}$ kg $\cdot$ m, and also note the scaling of the $y$ axis.}
\label{fig:C2H6_Y}
\end{figure}

For all of the simulations above studying the pseudoscalar $\mathcal{H}$, it is important to note that the results can in principal depend strongly on the axis chosen.
In Fig.~\ref{fig:C2H6_X} and Fig.~\ref{fig:C2H6_Y}, we plot results for ethane where we rotate the direction of this axis for the helicity calculation. In the main text, the direction of helicity calculation is defined along $z$ axis, taken along the C-C bond. The $x$ axis is defined parallel to the line segment connecting two hydrogen atoms connected to the same carbon atom, while the $y$ axis is defined to be perpendicular to the $x$ and $z$ axes such that the coordinate system is right handed. We rotate the axis of helicity calculation from $z$ axis to an axis in $x$-$z$ or $y$-$z$ plane, e.g. rotate it around $x$ or $y$ axis. We find that, for ethane, the results are not that sensitive to the choice of axis as far as the overall trend remains. We also do the same calculation for glycine, as the $z$-axis is defined along the C-C bond, the $x$-axis is defined to be perpendicular to the plane of C-C and C=O bond while the $y$-axis again is defined to be perpendicular to $x$ and $z$ axis. Here, we find that the results are extremely sensitive to the choice of axis such that, for a 90$^\circ$ angle rotation in Fig. \ref{fig:C2H5O2N_Y}c, the high and low frequencies flip helicity. This may be explained by the fact that glycine is much less symmetric than ethane. Clearly, one must be more careful in interpreting these momentum pseudoscalar calculations insomuch as the the results depend on the choice of axis.
\begin{figure}[hbt!]
\includegraphics[width=240pt]{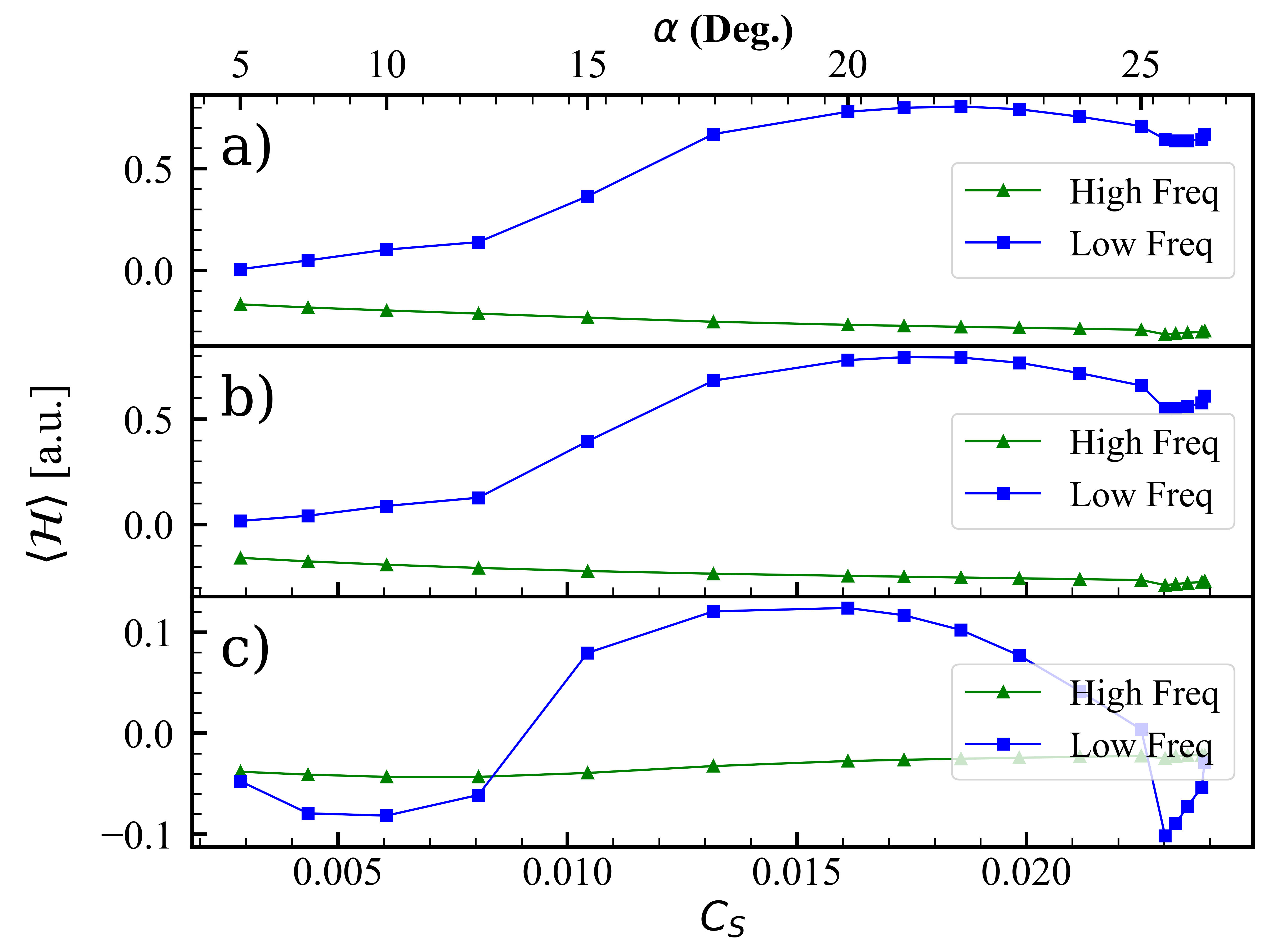}
\caption{ Rotating the axis of the momentum pseudoscalar of glycine around $x$ axis by (a) $30^\circ$ (b) $60^\circ$ (c) $90^\circ$ (Originally along $z$ axis) Helicity is reported in an arbitrary unit that is equivalent to $8.8 \times 10^{-38}$ kg $\cdot$ m, and also note the scaling of the $y$ axis.}
\label{fig:C2H5O2N_X}
\end{figure}
\begin{figure}[hbt!]
\includegraphics[width=240pt]{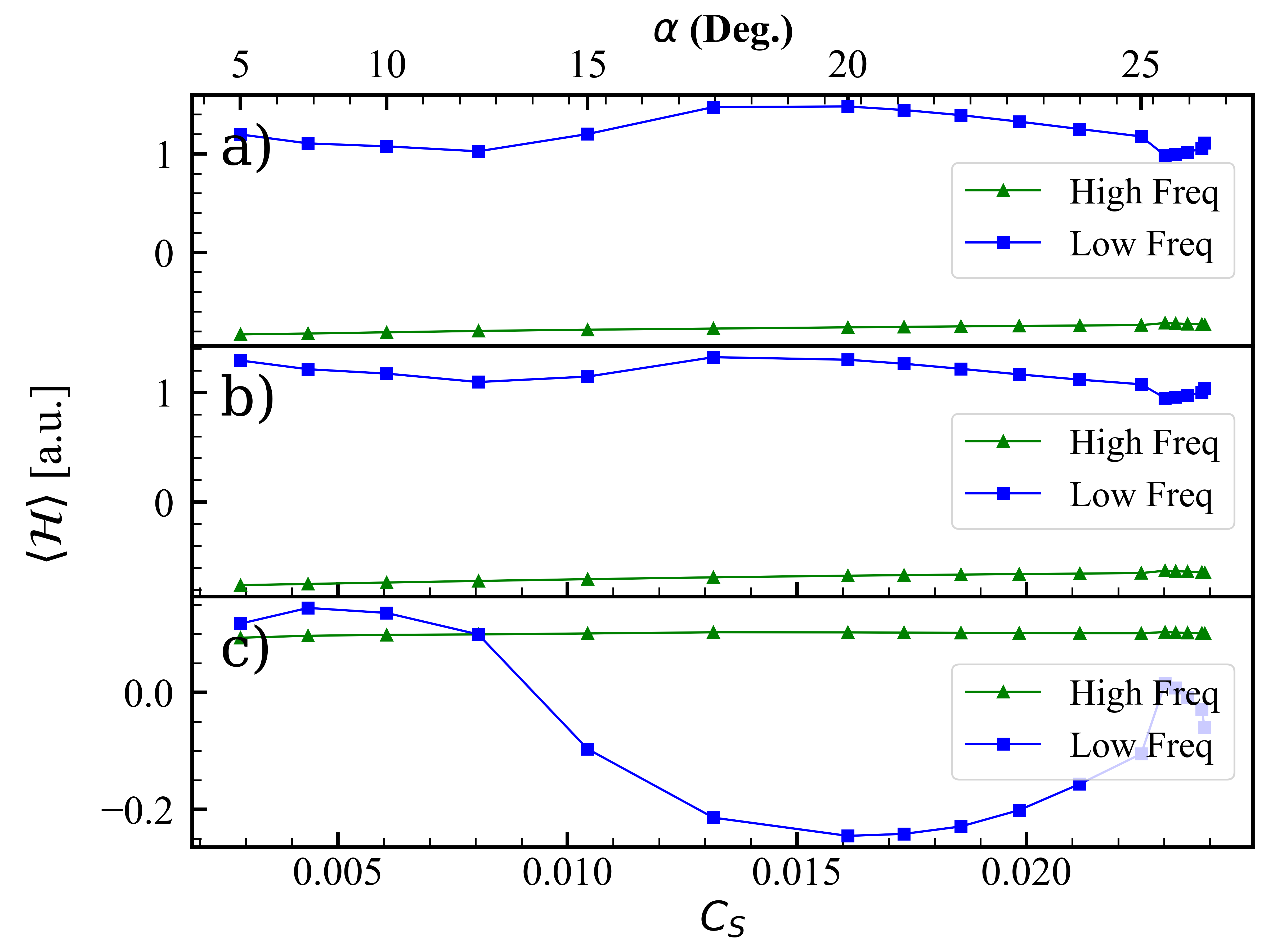}
\caption{ Rotating the axis of the momentum pseudoscalar of glycine around $y$ axis by (a) $30^\circ$ (b) $60^\circ$ (c) $90^\circ$ (Originally along $z$ axis) Helicity is reported in an arbitrary unit that is equivalent to $8.8 \times 10^{-38}$ kg $\cdot$ m, and also note the scaling of the $y$ axis.}
\label{fig:C2H5O2N_Y}
\end{figure}

\section{Quadratic expansion of $C_S$ under small twist angle}
Assume that the twisting procedure twists along the a specified dihedral while keeping the rest of the structure rigid. If the original conformation is achiral we have
\begin{equation}
    \begin{aligned}
        C_S=&\frac{1}{Z}\sum_i \left|\hat{\sigma}\mathbf{r_{\hat{p}(i)}}-\mathbf{r_i}   \right|^2\\
        =&0
    \end{aligned}
\end{equation}
where this alternate form of the CCM (in which $\hat{p}$ is the permutation operator and $Z$ is a normalization factor) is discussed in Ref. \citen{avnir1995CCM}. The equality to zero in Eq. S1, implied by the assumed achirality of the conformation, requires that $\hat{\sigma}\mathbf{r_{\mathit{\hat{p}(i)}}}=\mathbf{r_i}$ for every atom $i$.

If we twist the structure, it is clear that every atom in the structure is rotated around the twisting axis with angle $\alpha$. One subtlety is that the displacement of coordinates are mass-weighted, which means that the rotation angle of each atom $j$ must be scaled to an effective rotation angle $\theta_j=c_j \alpha$, where $c_j$ is determined by the relative mass of the atom $j$.

We could write the new $C_S$ of the twisted structure as
\begin{equation}
    \begin{aligned}
        C_S(\alpha)&=\frac{1}{Z}\sum_i \left|\hat{\sigma}\left[\mathbb{R}(\theta_{\hat{p}(i)})\mathbf{r_{\mathit{\hat{p}(i)}}}\right] - \mathbb{R}(\theta_i)\mathbf{r_i} \right|^2\\
        &=\frac{1}{Z}\sum_i \left|\mathbb{R}(-\theta_{\hat{p}(i)})\hat{\sigma}\mathbf{r_{\mathit{\hat{p}(i)}}} - \mathbb{R}(\theta_i)\mathbf{r_i} \right|^2\\
        &=\frac{1}{Z}\sum_i \left|\mathbb{R}(-\theta_{\hat{p}(i)})\mathbf{r_i} - \mathbb{R}(\theta_i)\mathbf{r_i} \right|^2.\\
    \end{aligned}
\end{equation}
where $\mathbb{R}$ is the rotation operator. Note that when $\alpha\sim0$, the rotation operator could be expressed as 
$\mathbb{R}(\mathrm{d}\theta)=\mathbb{I}+\mathrm{d}\theta\mathbb{A}$, where $\mathbb{A}$ is the infinitesimal rotation operator that specifies the axis of rotation. Thus, $C_S(\alpha)$ can be expressed as
\begin{equation}
    \begin{aligned}
        C_S(\alpha)&=\frac{1}{Z}\sum_i \left|\mathbb{R}(-\theta_{\hat{p}(i)})\mathbf{r_i} - \mathbb{R}(\theta_i)\mathbf{r_i} \right|^2\\
        &=\frac{1}{Z}\sum_i \left|\left[\mathbb{I}+(-\theta_{\hat{p}(i)})\mathbb{A}\right]\mathbf{r_i} - \left[\mathbb{I}+(\theta_i)\mathbb{A}\right]\mathbf{r_i} \right|^2\\
        &=\frac{1}{Z}\sum_i \left|-(\theta_{\hat{p}(i)}+\theta_i)\mathbb{A}\mathbf{r_i} \right|^2\\
        &=\left(\frac{1}{Z}\sum_i (c_{\hat{p}(i)}+c_i)^2\left|\mathbb{A}\mathbf{r_i} \right|^2\right)\alpha^2.\\
    \end{aligned}
\end{equation}
Since the factor inside the parentheses is constant, we have shown that $C_S \propto \alpha^2$. This result, that the CCM is quadratic in a small rotational deviation from an achiral structure, is expected to be general and applies also, for instance, to the 4-site model examined in Ref. \citen{ethan2024benchmark}.
\bibliography{reference}

\begin{thebibliography}{26}%
\makeatletter
\providecommand \@ifxundefined [1]{%
 \@ifx{#1\undefined}
}%
\providecommand \@ifnum [1]{%
 \ifnum #1\expandafter \@firstoftwo
 \else \expandafter \@secondoftwo
 \fi
}%
\providecommand \@ifx [1]{%
 \ifx #1\expandafter \@firstoftwo
 \else \expandafter \@secondoftwo
 \fi
}%
\providecommand \natexlab [1]{#1}%
\providecommand \enquote  [1]{``#1''}%
\providecommand \bibnamefont  [1]{#1}%
\providecommand \bibfnamefont [1]{#1}%
\providecommand \citenamefont [1]{#1}%
\providecommand \href@noop [0]{\@secondoftwo}%
\providecommand \href [0]{\begingroup \@sanitize@url \@href}%
\providecommand \@href[1]{\@@startlink{#1}\@@href}%
\providecommand \@@href[1]{\endgroup#1\@@endlink}%
\providecommand \@sanitize@url [0]{\catcode `\\12\catcode `\$12\catcode `\&12\catcode `\#12\catcode `\^12\catcode `\_12\catcode `\%12\relax}%
\providecommand \@@startlink[1]{}%
\providecommand \@@endlink[0]{}%
\providecommand \url  [0]{\begingroup\@sanitize@url \@url }%
\providecommand \@url [1]{\endgroup\@href {#1}{\urlprefix }}%
\providecommand \urlprefix  [0]{URL }%
\providecommand \Eprint [0]{\href }%
\providecommand \doibase [0]{http://dx.doi.org/}%
\providecommand \selectlanguage [0]{\@gobble}%
\providecommand \bibinfo  [0]{\@secondoftwo}%
\providecommand \bibfield  [0]{\@secondoftwo}%
\providecommand \translation [1]{[#1]}%
\providecommand \BibitemOpen [0]{}%
\providecommand \bibitemStop [0]{}%
\providecommand \bibitemNoStop [0]{.\EOS\space}%
\providecommand \EOS [0]{\spacefactor3000\relax}%
\providecommand \BibitemShut  [1]{\csname bibitem#1\endcsname}%
\let\auto@bib@innerbib\@empty
%</preamble>
\bibitem [{\citenamefont {Hicks}(2002)}]{hicks2002chirality}%
  \BibitemOpen
  \bibfield  {author} {\bibinfo {author} {\bibfnamefont {J.}~\bibnamefont {Hicks}},\ }\href {https://books.google.com/books?id=4RvwAAAAMAAJ} {\emph {\bibinfo {title} {Chirality: Physical Chemistry}}},\ ACS symposium series\ (\bibinfo  {publisher} {American Chemical Society},\ \bibinfo {year} {2002})\BibitemShut {NoStop}%
\bibitem [{\citenamefont {Fransson}(2019)}]{fransson2019theory}%
  \BibitemOpen
  \bibfield  {author} {\bibinfo {author} {\bibfnamefont {J.}~\bibnamefont {Fransson}},\ }\bibfield  {title} {\enquote {\bibinfo {title} {Chirality-induced spin selectivity: The role of electron correlations},}\ }\href@noop {} {\bibfield  {journal} {\bibinfo  {journal} {The journal of physical chemistry letters}\ }\textbf {\bibinfo {volume} {10}},\ \bibinfo {pages} {7126--7132} (\bibinfo {year} {2019})}\BibitemShut {NoStop}%
\bibitem [{\citenamefont {Wan}\ \emph {et~al.}(2023)\citenamefont {Wan}, \citenamefont {Liu}, \citenamefont {Fuchter},\ and\ \citenamefont {Yan}}]{yan2023locking}%
  \BibitemOpen
  \bibfield  {author} {\bibinfo {author} {\bibfnamefont {L.}~\bibnamefont {Wan}}, \bibinfo {author} {\bibfnamefont {Y.}~\bibnamefont {Liu}}, \bibinfo {author} {\bibfnamefont {M.~J.}\ \bibnamefont {Fuchter}}, \ and\ \bibinfo {author} {\bibfnamefont {B.}~\bibnamefont {Yan}},\ }\bibfield  {title} {\enquote {\bibinfo {title} {Anomalous circularly polarized light emission in organic light-emitting diodes caused by orbital--momentum locking},}\ }\href@noop {} {\bibfield  {journal} {\bibinfo  {journal} {Nature Photonics}\ }\textbf {\bibinfo {volume} {17}},\ \bibinfo {pages} {193--199} (\bibinfo {year} {2023})}\BibitemShut {NoStop}%
\bibitem [{\citenamefont {Yang}\ \emph {et~al.}(2023)\citenamefont {Yang}, \citenamefont {Xiao}, \citenamefont {Robredo}, \citenamefont {Vergniory}, \citenamefont {Yan},\ and\ \citenamefont {Felser}}]{yan2023locking2}%
  \BibitemOpen
  \bibfield  {author} {\bibinfo {author} {\bibfnamefont {Q.}~\bibnamefont {Yang}}, \bibinfo {author} {\bibfnamefont {J.}~\bibnamefont {Xiao}}, \bibinfo {author} {\bibfnamefont {I.}~\bibnamefont {Robredo}}, \bibinfo {author} {\bibfnamefont {M.~G.}\ \bibnamefont {Vergniory}}, \bibinfo {author} {\bibfnamefont {B.}~\bibnamefont {Yan}}, \ and\ \bibinfo {author} {\bibfnamefont {C.}~\bibnamefont {Felser}},\ }\bibfield  {title} {\enquote {\bibinfo {title} {Monopole-like orbital-momentum locking and the induced orbital transport in topological chiral semimetals},}\ }\href@noop {} {\bibfield  {journal} {\bibinfo  {journal} {Proceedings of the National Academy of Sciences}\ }\textbf {\bibinfo {volume} {120}},\ \bibinfo {pages} {e2305541120} (\bibinfo {year} {2023})}\BibitemShut {NoStop}%
\bibitem [{\citenamefont {Yan}(2024)}]{yan2024lockingreview}%
  \BibitemOpen
  \bibfield  {author} {\bibinfo {author} {\bibfnamefont {B.}~\bibnamefont {Yan}},\ }\bibfield  {title} {\enquote {\bibinfo {title} {Structural chirality and electronic chirality in quantum materials},}\ }\href@noop {} {\bibfield  {journal} {\bibinfo  {journal} {Annual Review of Materials Research}\ }\textbf {\bibinfo {volume} {54}} (\bibinfo {year} {2024})}\BibitemShut {NoStop}%
\bibitem [{\citenamefont {Skourtis}\ \emph {et~al.}(2008)\citenamefont {Skourtis}, \citenamefont {Beratan}, \citenamefont {Naaman}, \citenamefont {Nitzan},\ and\ \citenamefont {Waldeck}}]{abe2008theory}%
  \BibitemOpen
  \bibfield  {author} {\bibinfo {author} {\bibfnamefont {S.~S.}\ \bibnamefont {Skourtis}}, \bibinfo {author} {\bibfnamefont {D.~N.}\ \bibnamefont {Beratan}}, \bibinfo {author} {\bibfnamefont {R.}~\bibnamefont {Naaman}}, \bibinfo {author} {\bibfnamefont {A.}~\bibnamefont {Nitzan}}, \ and\ \bibinfo {author} {\bibfnamefont {D.~H.}\ \bibnamefont {Waldeck}},\ }\bibfield  {title} {\enquote {\bibinfo {title} {Chiral control of electron transmission through molecules},}\ }\href@noop {} {\bibfield  {journal} {\bibinfo  {journal} {Physical review letters}\ }\textbf {\bibinfo {volume} {101}},\ \bibinfo {pages} {238103} (\bibinfo {year} {2008})}\BibitemShut {NoStop}%
\bibitem [{\citenamefont {Gersten}, \citenamefont {Kaasbjerg},\ and\ \citenamefont {Nitzan}(2013)}]{abe2013theory}%
  \BibitemOpen
  \bibfield  {author} {\bibinfo {author} {\bibfnamefont {J.}~\bibnamefont {Gersten}}, \bibinfo {author} {\bibfnamefont {K.}~\bibnamefont {Kaasbjerg}}, \ and\ \bibinfo {author} {\bibfnamefont {A.}~\bibnamefont {Nitzan}},\ }\bibfield  {title} {\enquote {\bibinfo {title} {Induced spin filtering in electron transmission through chiral molecular layers adsorbed on metals with strong spin-orbit coupling},}\ }\href@noop {} {\bibfield  {journal} {\bibinfo  {journal} {The Journal of chemical physics}\ }\textbf {\bibinfo {volume} {139}} (\bibinfo {year} {2013})}\BibitemShut {NoStop}%
\bibitem [{\citenamefont {Naaman}\ and\ \citenamefont {Waldeck}(2012)}]{naaman2012exp}%
  \BibitemOpen
  \bibfield  {author} {\bibinfo {author} {\bibfnamefont {R.}~\bibnamefont {Naaman}}\ and\ \bibinfo {author} {\bibfnamefont {D.~H.}\ \bibnamefont {Waldeck}},\ }\bibfield  {title} {\enquote {\bibinfo {title} {Chiral-induced spin selectivity effect},}\ }\href@noop {} {\bibfield  {journal} {\bibinfo  {journal} {The journal of physical chemistry letters}\ }\textbf {\bibinfo {volume} {3}},\ \bibinfo {pages} {2178--2187} (\bibinfo {year} {2012})}\BibitemShut {NoStop}%
\bibitem [{\citenamefont {Naaman}, \citenamefont {Paltiel},\ and\ \citenamefont {Waldeck}(2020)}]{naaman2020exp}%
  \BibitemOpen
  \bibfield  {author} {\bibinfo {author} {\bibfnamefont {R.}~\bibnamefont {Naaman}}, \bibinfo {author} {\bibfnamefont {Y.}~\bibnamefont {Paltiel}}, \ and\ \bibinfo {author} {\bibfnamefont {D.~H.}\ \bibnamefont {Waldeck}},\ }\bibfield  {title} {\enquote {\bibinfo {title} {Chiral induced spin selectivity gives a new twist on spin-control in chemistry},}\ }\href@noop {} {\bibfield  {journal} {\bibinfo  {journal} {Accounts of Chemical Research}\ }\textbf {\bibinfo {volume} {53}},\ \bibinfo {pages} {2659--2667} (\bibinfo {year} {2020})}\BibitemShut {NoStop}%
\bibitem [{\citenamefont {Bloom}\ \emph {et~al.}(2024)\citenamefont {Bloom}, \citenamefont {Paltiel}, \citenamefont {Naaman},\ and\ \citenamefont {Waldeck}}]{waldek2024review}%
  \BibitemOpen
  \bibfield  {author} {\bibinfo {author} {\bibfnamefont {B.~P.}\ \bibnamefont {Bloom}}, \bibinfo {author} {\bibfnamefont {Y.}~\bibnamefont {Paltiel}}, \bibinfo {author} {\bibfnamefont {R.}~\bibnamefont {Naaman}}, \ and\ \bibinfo {author} {\bibfnamefont {D.~H.}\ \bibnamefont {Waldeck}},\ }\bibfield  {title} {\enquote {\bibinfo {title} {Chiral induced spin selectivity},}\ }\href@noop {} {\bibfield  {journal} {\bibinfo  {journal} {Chemical Reviews}\ }\textbf {\bibinfo {volume} {124}},\ \bibinfo {pages} {1950--1991} (\bibinfo {year} {2024})}\BibitemShut {NoStop}%
\bibitem [{\citenamefont {Zhang}\ and\ \citenamefont {Niu}(2014{\natexlab{a}})}]{zhang2014theory}%
  \BibitemOpen
  \bibfield  {author} {\bibinfo {author} {\bibfnamefont {L.}~\bibnamefont {Zhang}}\ and\ \bibinfo {author} {\bibfnamefont {Q.}~\bibnamefont {Niu}},\ }\bibfield  {title} {\enquote {\bibinfo {title} {Angular momentum of phonons and the einstein--de haas effect},}\ }\href@noop {} {\bibfield  {journal} {\bibinfo  {journal} {Physical Review Letters}\ }\textbf {\bibinfo {volume} {112}},\ \bibinfo {pages} {085503} (\bibinfo {year} {2014}{\natexlab{a}})}\BibitemShut {NoStop}%
\bibitem [{\citenamefont {Zhang}\ and\ \citenamefont {Niu}(2015)}]{zhang2015theory}%
  \BibitemOpen
  \bibfield  {author} {\bibinfo {author} {\bibfnamefont {L.}~\bibnamefont {Zhang}}\ and\ \bibinfo {author} {\bibfnamefont {Q.}~\bibnamefont {Niu}},\ }\bibfield  {title} {\enquote {\bibinfo {title} {Chiral phonons at high-symmetry points in monolayer hexagonal lattices},}\ }\href@noop {} {\bibfield  {journal} {\bibinfo  {journal} {Physical review letters}\ }\textbf {\bibinfo {volume} {115}},\ \bibinfo {pages} {115502} (\bibinfo {year} {2015})}\BibitemShut {NoStop}%
\bibitem [{\citenamefont {Chen}\ \emph {et~al.}(2019)\citenamefont {Chen}, \citenamefont {Wu}, \citenamefont {Yang}, \citenamefont {Li},\ and\ \citenamefont {Zhang}}]{zhang2019theory}%
  \BibitemOpen
  \bibfield  {author} {\bibinfo {author} {\bibfnamefont {H.}~\bibnamefont {Chen}}, \bibinfo {author} {\bibfnamefont {W.}~\bibnamefont {Wu}}, \bibinfo {author} {\bibfnamefont {S.~A.}\ \bibnamefont {Yang}}, \bibinfo {author} {\bibfnamefont {X.}~\bibnamefont {Li}}, \ and\ \bibinfo {author} {\bibfnamefont {L.}~\bibnamefont {Zhang}},\ }\bibfield  {title} {\enquote {\bibinfo {title} {Chiral phonons in kagome lattices},}\ }\href@noop {} {\bibfield  {journal} {\bibinfo  {journal} {Physical Review B}\ }\textbf {\bibinfo {volume} {100}},\ \bibinfo {pages} {094303} (\bibinfo {year} {2019})}\BibitemShut {NoStop}%
\bibitem [{\citenamefont {Zhu}\ \emph {et~al.}(2018)\citenamefont {Zhu}, \citenamefont {Yi}, \citenamefont {Li}, \citenamefont {Xiao}, \citenamefont {Zhang}, \citenamefont {Yang}, \citenamefont {Kaindl}, \citenamefont {Li}, \citenamefont {Wang},\ and\ \citenamefont {Zhang}}]{zhu2018phononexp}%
  \BibitemOpen
  \bibfield  {author} {\bibinfo {author} {\bibfnamefont {H.}~\bibnamefont {Zhu}}, \bibinfo {author} {\bibfnamefont {J.}~\bibnamefont {Yi}}, \bibinfo {author} {\bibfnamefont {M.-Y.}\ \bibnamefont {Li}}, \bibinfo {author} {\bibfnamefont {J.}~\bibnamefont {Xiao}}, \bibinfo {author} {\bibfnamefont {L.}~\bibnamefont {Zhang}}, \bibinfo {author} {\bibfnamefont {C.-W.}\ \bibnamefont {Yang}}, \bibinfo {author} {\bibfnamefont {R.~A.}\ \bibnamefont {Kaindl}}, \bibinfo {author} {\bibfnamefont {L.-J.}\ \bibnamefont {Li}}, \bibinfo {author} {\bibfnamefont {Y.}~\bibnamefont {Wang}}, \ and\ \bibinfo {author} {\bibfnamefont {X.}~\bibnamefont {Zhang}},\ }\bibfield  {title} {\enquote {\bibinfo {title} {Observation of chiral phonons},}\ }\href@noop {} {\bibfield  {journal} {\bibinfo  {journal} {Science}\ }\textbf {\bibinfo {volume} {359}},\ \bibinfo {pages} {579--582} (\bibinfo {year} {2018})}\BibitemShut {NoStop}%
\bibitem [{\citenamefont {Chen}\ \emph {et~al.}(2018)\citenamefont {Chen}, \citenamefont {Zhang}, \citenamefont {Niu},\ and\ \citenamefont {Zhang}}]{chen2018review}%
  \BibitemOpen
  \bibfield  {author} {\bibinfo {author} {\bibfnamefont {H.}~\bibnamefont {Chen}}, \bibinfo {author} {\bibfnamefont {W.}~\bibnamefont {Zhang}}, \bibinfo {author} {\bibfnamefont {Q.}~\bibnamefont {Niu}}, \ and\ \bibinfo {author} {\bibfnamefont {L.}~\bibnamefont {Zhang}},\ }\bibfield  {title} {\enquote {\bibinfo {title} {Chiral phonons in two-dimensional materials},}\ }\href@noop {} {\bibfield  {journal} {\bibinfo  {journal} {2D Materials}\ }\textbf {\bibinfo {volume} {6}},\ \bibinfo {pages} {012002} (\bibinfo {year} {2018})}\BibitemShut {NoStop}%
\bibitem [{\citenamefont {Chen}\ \emph {et~al.}(2022)\citenamefont {Chen}, \citenamefont {Wu}, \citenamefont {Zhu}, \citenamefont {Yang}, \citenamefont {Gong}, \citenamefont {Gao}, \citenamefont {Yang},\ and\ \citenamefont {Zhang}}]{chen2022diodetheory}%
  \BibitemOpen
  \bibfield  {author} {\bibinfo {author} {\bibfnamefont {H.}~\bibnamefont {Chen}}, \bibinfo {author} {\bibfnamefont {W.}~\bibnamefont {Wu}}, \bibinfo {author} {\bibfnamefont {J.}~\bibnamefont {Zhu}}, \bibinfo {author} {\bibfnamefont {Z.}~\bibnamefont {Yang}}, \bibinfo {author} {\bibfnamefont {W.}~\bibnamefont {Gong}}, \bibinfo {author} {\bibfnamefont {W.}~\bibnamefont {Gao}}, \bibinfo {author} {\bibfnamefont {S.~A.}\ \bibnamefont {Yang}}, \ and\ \bibinfo {author} {\bibfnamefont {L.}~\bibnamefont {Zhang}},\ }\bibfield  {title} {\enquote {\bibinfo {title} {Chiral phonon diode effect in chiral crystals},}\ }\href@noop {} {\bibfield  {journal} {\bibinfo  {journal} {Nano Letters}\ }\textbf {\bibinfo {volume} {22}},\ \bibinfo {pages} {1688--1693} (\bibinfo {year} {2022})}\BibitemShut {NoStop}%
\bibitem [{\citenamefont {Harris}, \citenamefont {Kamien},\ and\ \citenamefont {Lubensky}(1999)}]{Kamien}%
  \BibitemOpen
  \bibfield  {author} {\bibinfo {author} {\bibfnamefont {A.~B.}\ \bibnamefont {Harris}}, \bibinfo {author} {\bibfnamefont {R.~D.}\ \bibnamefont {Kamien}}, \ and\ \bibinfo {author} {\bibfnamefont {T.~C.}\ \bibnamefont {Lubensky}},\ }\bibfield  {title} {\enquote {\bibinfo {title} {Molecular chirality and chiral parameters},}\ }\href {\doibase 10.1103/RevModPhys.71.1745} {\bibfield  {journal} {\bibinfo  {journal} {Rev. Mod. Phys.}\ }\textbf {\bibinfo {volume} {71}},\ \bibinfo {pages} {1745--1757} (\bibinfo {year} {1999})}\BibitemShut {NoStop}%
\bibitem [{\citenamefont {Abraham}\ and\ \citenamefont {Nitzan}(2024{\natexlab{a}})}]{ethan2023quantifying}%
  \BibitemOpen
  \bibfield  {author} {\bibinfo {author} {\bibfnamefont {E.}~\bibnamefont {Abraham}}\ and\ \bibinfo {author} {\bibfnamefont {A.}~\bibnamefont {Nitzan}},\ }\bibfield  {title} {\enquote {\bibinfo {title} {Quantifying the chirality of vibrational modes in helical molecular chains},}\ }\href {\doibase 10.1103/PhysRevLett.133.268001} {\bibfield  {journal} {\bibinfo  {journal} {Phys. Rev. Lett.}\ }\textbf {\bibinfo {volume} {133}},\ \bibinfo {pages} {268001} (\bibinfo {year} {2024}{\natexlab{a}})}\BibitemShut {NoStop}%
\bibitem [{\citenamefont {Abraham}\ and\ \citenamefont {Nitzan}(2024{\natexlab{b}})}]{ethan2024benchmark}%
  \BibitemOpen
  \bibfield  {author} {\bibinfo {author} {\bibfnamefont {E.}~\bibnamefont {Abraham}}\ and\ \bibinfo {author} {\bibfnamefont {A.}~\bibnamefont {Nitzan}},\ }\bibfield  {title} {\enquote {\bibinfo {title} {Molecular chirality quantification: Tools and benchmarks},}\ }\href@noop {} {\bibfield  {journal} {\bibinfo  {journal} {The Journal of Chemical Physics}\ }\textbf {\bibinfo {volume} {160}} (\bibinfo {year} {2024}{\natexlab{b}})}\BibitemShut {NoStop}%
\bibitem [{\citenamefont {Zabrodsky}\ and\ \citenamefont {Avnir}(1995)}]{avnir1995CCM}%
  \BibitemOpen
  \bibfield  {author} {\bibinfo {author} {\bibfnamefont {H.}~\bibnamefont {Zabrodsky}}\ and\ \bibinfo {author} {\bibfnamefont {D.}~\bibnamefont {Avnir}},\ }\bibfield  {title} {\enquote {\bibinfo {title} {Continuous symmetry measures. 4. chirality},}\ }\href@noop {} {\bibfield  {journal} {\bibinfo  {journal} {Journal of the American Chemical Society}\ }\textbf {\bibinfo {volume} {117}},\ \bibinfo {pages} {462--473} (\bibinfo {year} {1995})}\BibitemShut {NoStop}%
\bibitem [{\citenamefont {Alvarez}, \citenamefont {Alemany},\ and\ \citenamefont {Avnir}(2005)}]{avnir2005CCM}%
  \BibitemOpen
  \bibfield  {author} {\bibinfo {author} {\bibfnamefont {S.}~\bibnamefont {Alvarez}}, \bibinfo {author} {\bibfnamefont {P.}~\bibnamefont {Alemany}}, \ and\ \bibinfo {author} {\bibfnamefont {D.}~\bibnamefont {Avnir}},\ }\bibfield  {title} {\enquote {\bibinfo {title} {Continuous chirality measures in transition metal chemistry},}\ }\href@noop {} {\bibfield  {journal} {\bibinfo  {journal} {Chemical Society Reviews}\ }\textbf {\bibinfo {volume} {34}},\ \bibinfo {pages} {313--326} (\bibinfo {year} {2005})}\BibitemShut {NoStop}%
\bibitem [{\citenamefont {Pinsky}\ \emph {et~al.}(2008)\citenamefont {Pinsky}, \citenamefont {Dryzun}, \citenamefont {Casanova}, \citenamefont {Alemany},\ and\ \citenamefont {Avnir}}]{avnir2008CCM}%
  \BibitemOpen
  \bibfield  {author} {\bibinfo {author} {\bibfnamefont {M.}~\bibnamefont {Pinsky}}, \bibinfo {author} {\bibfnamefont {C.}~\bibnamefont {Dryzun}}, \bibinfo {author} {\bibfnamefont {D.}~\bibnamefont {Casanova}}, \bibinfo {author} {\bibfnamefont {P.}~\bibnamefont {Alemany}}, \ and\ \bibinfo {author} {\bibfnamefont {D.}~\bibnamefont {Avnir}},\ }\bibfield  {title} {\enquote {\bibinfo {title} {Analytical methods for calculating continuous symmetry measures and the chirality measure},}\ }\href@noop {} {\bibfield  {journal} {\bibinfo  {journal} {Journal of computational chemistry}\ }\textbf {\bibinfo {volume} {29}},\ \bibinfo {pages} {2712--2721} (\bibinfo {year} {2008})}\BibitemShut {NoStop}%
\bibitem [{\citenamefont {Abraham}\ \emph {et~al.}(2023)\citenamefont {Abraham}, \citenamefont {Dinpajooh}, \citenamefont {Climent},\ and\ \citenamefont {Nitzan}}]{ethan2023chain}%
  \BibitemOpen
  \bibfield  {author} {\bibinfo {author} {\bibfnamefont {E.}~\bibnamefont {Abraham}}, \bibinfo {author} {\bibfnamefont {M.}~\bibnamefont {Dinpajooh}}, \bibinfo {author} {\bibfnamefont {C.}~\bibnamefont {Climent}}, \ and\ \bibinfo {author} {\bibfnamefont {A.}~\bibnamefont {Nitzan}},\ }\bibfield  {title} {\enquote {\bibinfo {title} {Heat transport with a twist},}\ }\href@noop {} {\bibfield  {journal} {\bibinfo  {journal} {The Journal of Chemical Physics}\ }\textbf {\bibinfo {volume} {159}} (\bibinfo {year} {2023})}\BibitemShut {NoStop}%
\bibitem [{Note1()}]{Note1}%
  \BibitemOpen
  \bibinfo {note} {In some published studies the permutations $\protect \mathbf {Q} \rightarrow \protect \mathbf {P}$ are avoided altogether so that $\protect \mathbf {P}$ is replaced by $\protect \mathbf {Q}$ in Eq.~\ref {equation:CCM}. This so called convention of trivial permutations used to deal with large systems where the complexity of the calculation rapidly increases with the number of identical atoms. Here, we do not restrict ourselves to such conventions, as our system is not that large, allowing us to find the best permutation. Some other considerations associated with the applications of permutations are presented in Appendix A.}\BibitemShut {Stop}%
\bibitem [{\citenamefont {Zhang}\ and\ \citenamefont {Niu}(2014{\natexlab{b}})}]{phon_ang_mom}%
  \BibitemOpen
  \bibfield  {author} {\bibinfo {author} {\bibfnamefont {L.}~\bibnamefont {Zhang}}\ and\ \bibinfo {author} {\bibfnamefont {Q.}~\bibnamefont {Niu}},\ }\bibfield  {title} {\enquote {\bibinfo {title} {Angular momentum of phonons and the einstein--de haas effect},}\ }\href {\doibase 10.1103/PhysRevLett.112.085503} {\bibfield  {journal} {\bibinfo  {journal} {Phys. Rev. Lett.}\ }\textbf {\bibinfo {volume} {112}},\ \bibinfo {pages} {085503} (\bibinfo {year} {2014}{\natexlab{b}})}\BibitemShut {NoStop}%
\bibitem [{\citenamefont {Hamada}\ \emph {et~al.}(2018)\citenamefont {Hamada}, \citenamefont {Minamitani}, \citenamefont {Hirayama},\ and\ \citenamefont {Murakami}}]{therm_angular}%
  \BibitemOpen
  \bibfield  {author} {\bibinfo {author} {\bibfnamefont {M.}~\bibnamefont {Hamada}}, \bibinfo {author} {\bibfnamefont {E.}~\bibnamefont {Minamitani}}, \bibinfo {author} {\bibfnamefont {M.}~\bibnamefont {Hirayama}}, \ and\ \bibinfo {author} {\bibfnamefont {S.}~\bibnamefont {Murakami}},\ }\bibfield  {title} {\enquote {\bibinfo {title} {Phonon angular momentum induced by the temperature gradient},}\ }\href {\doibase 10.1103/PhysRevLett.121.175301} {\bibfield  {journal} {\bibinfo  {journal} {Phys. Rev. Lett.}\ }\textbf {\bibinfo {volume} {121}},\ \bibinfo {pages} {175301} (\bibinfo {year} {2018})}\BibitemShut {NoStop}%
\end{thebibliography}%
\end{document}